\documentclass[12pt]{article}%
\usepackage{amsmath}
\usepackage{amsfonts}
\usepackage{amssymb}
\usepackage{graphicx}%
\providecommand{\U}[1]{\protect\rule{.1in}{.1in}}

\begin{document}

\title{Oscillations of the magnetic polarization in a Kondo impurity at finite
magnetic fields.}
\author{Gerd Bergmann\\Department of Physics\\University of Southern California\\Los Angeles, California 90089-0484\\e-mail: bergmann@usc.edu}
\date{\today }
\maketitle

\begin{abstract}
The electronic properties of a Kondo impurity are investigated in a magnetic
field using linear response theory. The distribution of electrical charge and
magnetic polarization are calculated in real space. The (small) magnetic field
does not change the charge distribution. However, it unmasks the Kondo cloud.
The (equal) weight of the d-electron components with their magnetic moment up
and down is shifted and the compensating s-electron clouds don't cancel any
longer (a requirement for an experimental detection of the Kondo cloud). In
addition to the net magnetic polarization of the conduction electrons an
oscillating magnetic polarization with a period of half the Fermi wave length
is observed. However, this oscillating magnetic polarization does not show the
long range behavior of Rudermann-Kittel-Kasuya-Yosida oscillations because the
oscillations don't extend beyond the Kondo radius. They represent an internal
electronic structure of the Kondo impurity in a magnetic field.

 \newpage

\end{abstract}

\section{Introduction}

The properties of magnetic impurities in a metal is one of the most
intensively studied problems in solid state physics. The magnetic impurity
combines two fascinating phenomena, (i) the formation of a magnetic moment
(\textbf{MM}) in a host \cite{F28}, \cite{A31} due to the Coulomb interaction
and (ii) the formation of a non-magnetic singlet state at low temperatures
\cite{K8}, \cite{Y2}, \cite{V7}, \cite{S77}, \cite{D44}, \cite{H23},
\cite{M20}, \cite{A36}, \cite{G24}, \cite{C8}, \cite{H20}. The electronic
structure of this singlet or Kondo ground state has been investigated using a
large number of sophisticated methods, for example: scaling \cite{A51},
renormalization \cite{W18}, \cite{F30}, \cite{K58}, \cite{K59} Fermi-liquid
theory \cite{N14}, \cite{N5}, slave-bosons (see for example \cite{N7}),
large-spin limit \cite{G19}, \cite{B103}, local moment approach \cite{L57},
\cite{L56}, Bethe ansatz \cite{W12}, \cite{A50}, \cite{S29}. The author has
added in recent years a new approach, the FAIR method, (FAIR for
\textbf{F}riedel \textbf{A}rtificially \textbf{I}nserted \textbf{R}esonance)
which yields a very compact approximate ground state of the Friedel-Anderson
and the Kondo impurity \cite{B151}, \cite{B152}, \cite{B153}.

One of the most controversial aspects of the Kondo ground state is the
so-called Kondo cloud within the radius $r_{K}$ where $r_{K}$ is called the
Kondo length
\begin{equation}
r_{K}=\frac{\hbar v_{F}}{k_{B}T_{K}} \label{K_xi}%
\end{equation}
($k_{B}T_{K}$ =$\varepsilon_{K}$= Kondo energy, $v_{F}$ = Fermi velocity of
the conduction electrons).

The idea is to divide the ground state $\Psi_{K}$ of a Kondo impurity into two
parts with opposite d-spins. The question is whether the conduction electrons
compensate the magnetic moment of the d-electron forming of a Kondo cloud in
each part. This question has been highly controversial over the last 30 years
(for references see for example \cite{G55}).

Recently the author \cite{B177} calculated the internal electronic structure
of a Kondo impurity in real space using the FAIR solution of the Kondo ground
state. This solution consists of only four Slater states and is well suited to
calculate spatial properties which are difficult to obtain with other methods.
The calculation showed that indeed a Kondo cloud of s-electrons compensates
perfectly the magnetic moment of the d-electrons in the Kondo ground state.

However, to study the polarization Kondo cloud experimentally one needs a
finite magnetic field because in zero magnetic field the polarization clouds
of the magnetic components cancel. The magnetic field should be small enough
so that the Kondo state is neither destroyed nor dramatically changed by the
field. Therefore, in this paper I will use linear response theory to calculate
the electronic structure of a Kondo impurity in a small magnetic field.

The magnetic properties, such as susceptibility, of the Kondo ground state in
a magnetic field have been studied theoretically in a number of papers
\cite{O36}, \cite{C31}, \cite{C32}, \cite{P45}. However, to the best of my
knowledge this is the first investigation of the effect of a magnetic field on
the electronic structure in real space. The number of real space calculations
in the Kondo ground state is still rather limited. Recently Affleck, Borda and
Saleur \cite{A83} calculated the Friedel oscillations in the vicinity of a
Kondo impurity by means of NRG calculations (\textbf{N}umerical \textbf{R}%
enormalization \textbf{G}roup). The author repeated their calculation with the
FAIR method and found good agreement with the NRG approach.

\section{Theoretical Background}

\subsection{The FAIR approach}

The basic idea of the FAIR method can be best explained for a Friedel
resonance with the Hamiltonian
\[
H_{F}=\sum_{\nu=0}^{N-1}\varepsilon_{\nu}c_{\nu}^{\dag}c_{\nu}+E_{d}d^{\dag
}d+\sum_{\nu=0}^{N-1}V_{sd}(\nu)[d^{\dag}c_{\nu}+c_{\nu}^{\dag}d]
\]
It has been shown by the author \cite{B91}, \cite{B92} that the \textbf{exact}
$n$-particle ground state of a Friedel Hamiltonian can be expressed as by the
sum of two Slater states.
\begin{equation}
\Psi_{F}=\left(  Aa_{0}^{\dag}+Bd^{\dag}\right)
{\textstyle\prod\limits_{i=1}^{n-1}}
a_{i}^{\dag}\Phi_{0} \label{Psi_F}%
\end{equation}
The state $a_{0}^{\dag}=%
{\textstyle\sum}
\alpha_{0}^{\nu}c_{\nu}^{\dag}$ is a localized state which is composed of the
states $c_{\nu}^{\dag}$. The states $\left\{  a_{i}^{\dag}\right\}  $
represent an $N$-dimensional orthonormal basis representing the same Hilbert
space as the basis $\left\{  c_{\nu}^{\dag}\right\}  $. The state $a_{0}%
^{\dag}$ determines uniquely the full basis $\left\{  a_{i}^{\dag}\right\}  $.

In this basis the conduction electron Hamiltonian $H_{0}=\sum_{\nu=1}%
^{N\ }\varepsilon_{\nu}c_{\nu}^{\dag}c_{\nu}$ takes the form%
\[
H_{0}=%
{\textstyle\sum_{i=1}^{N-1}}
E_{i}a_{i}^{\dag}a_{i}+E_{0}a_{0}^{\dag}a_{i}+%
{\textstyle\sum_{i}}
V_{fr}\left(  i\right)  \left[  a_{0}^{\dag}a_{i}+a_{i}^{\dag}a_{0}\right]
\]
As one can see the structure of $H_{0}$ is identical to the structure of the
Friedel Hamiltonian. The state $a_{0}^{\dag}$ represents an artificially
inserted Friedel resonance state. Therefore I call $a_{0}^{\dag}$ a
"\textbf{F}riedel \textbf{A}rtificially \textbf{I}nserted \textbf{R}esonance"
state or \textbf{FAIR}-state. The use of the FAIR-states is at the heart of my
approach to the FA- and Kondo impurity problem. Therefore I call this approach
the \textbf{FAIR }method.

In the ground state $\Psi_{F}$ the $\left(  n-1\right)  $ lowest states
$a_{i}^{\dag}$ are occupied (starting at $i=1$). The states $a_{0}^{\dag}$ and
$d^{\dag}$ are mixed. Further details are reviewed in the appendix.

\subsection{The Kondo impurity in a magnetic field}

The Kondo Hamiltonian in zero magnetic field has the form
\begin{equation}
H=%
{\textstyle\sum_{\nu=0}^{N-1}}
{\textstyle\sum_{\alpha}}
\varepsilon_{\nu}c_{\nu,\alpha}^{\dag}c_{\nu,\alpha}+%
{\textstyle\sum_{\nu,\nu^{\prime}=0}^{N-1}}
2J_{\nu,\nu^{\prime}}%
{\textstyle\sum_{\alpha,\alpha^{\prime}}}
c_{\nu,\alpha}^{\dag}\mathbf{s}_{\alpha,\alpha^{\prime}}c_{\nu^{\prime}%
,\alpha^{\prime}}\cdot\mathbf{S} \label{H_K}%
\end{equation}
where $\mathbf{s}=\mathbf{\sigma/2}$ is half the Pauli matrix vector,
$\mathbf{S}$ is the impurity spin $1/2.$The conduction electron band is
expressed by discrete Wilson states $c_{\nu}^{\dag}$ which have a logarithmic
energy scale (see appendix). $J_{\nu,\nu^{\prime}}$ is the matrix element of
the exchange interaction between the Wilson states $c_{\nu}^{\dag}$ and
$c_{\nu^{\prime}}^{\dag}$ via the impurity.

In previous papers the author introduced a very compact (approximate) solution
for the Kondo Hamiltonian which consists of four Slater states. (The path to
this solution is briefly sketched in the appendix).%

\begin{equation}
\psi_{K}=Ba_{0\uparrow}^{\dag}d_{\downarrow}^{\dag}\left\vert \mathbf{0}%
_{a\uparrow}\mathbf{0}_{b\downarrow}\right\rangle +Bd_{\uparrow}^{\dag
}a_{0\downarrow}^{\dag}\left\vert \mathbf{0}_{b\uparrow}\mathbf{0}%
_{a\downarrow}\right\rangle +Cd_{\uparrow}^{\dag}b_{0\downarrow}^{\dag
}\left\vert \mathbf{0}_{a\uparrow}\mathbf{0}_{b\downarrow}\right\rangle
+Cb_{0\uparrow}^{\dag}d_{\downarrow}^{\dag}\left\vert \mathbf{0}_{b\uparrow
}\mathbf{0}_{a\downarrow}\right\rangle \label{Psi_K}%
\end{equation}
Because of the spin degeneracy of the electrons two FAIR states $a_{0}^{\dag}$
and $b_{0}^{\dag}$ and the corresponding bases $\left\{  a_{i}^{\dag}\right\}
$ and $\left\{  b_{i}^{\dag}\right\}  $ are needed. The many-electron state
$\left\vert \mathbf{0}_{a\uparrow}\mathbf{0}_{b\downarrow}\right\rangle $
($\left\vert \mathbf{0}_{b\uparrow}\mathbf{0}_{a\downarrow}\right\rangle $) is
defined as%
\[
\left\vert \mathbf{0}_{a\uparrow}\mathbf{0}_{b\downarrow}\right\rangle =%
{\textstyle\prod\limits_{j=1}^{n-1}}
a_{j\uparrow}^{\dag}%
{\textstyle\prod_{k=1}^{n-1}}
a_{k\downarrow}^{\dag}\left\vert \Phi_{0}\right\rangle
\]
The FAIR states $a_{0,\sigma}^{\dag}$ ($b_{0,\sigma}^{\dag}$) and the d states
$d_{-\sigma}^{\dag}$ form a bound two-electron state with negative energy.

The four Slater states are abbreviated as
\begin{equation}
\Psi_{K}=B\Psi_{ad}+B\Psi_{da}+C\Psi_{db}+C\Psi_{bd} \label{Psi_K'}%
\end{equation}
(The interpretaton of $\Psi_{ad}$ is that the MM-up electrons use the basis
$\left\{  a_{i\uparrow}^{\dag}\right\}  $ and the state $a_{0\uparrow}^{\dag}$
is occupied ($d_{\uparrow}$ empty) while the MM-down electrons use the other
basis $\left\{  b_{i\downarrow}^{\dag}\right\}  $ and the state $d_{\downarrow
}^{\dag}$ is occupied ($b_{0\downarrow}^{\dag}$ empty). Since the ansatz for
the Kondo ground state consists of four Slater states its secular matrix has
the dimension four. One may consider the four Slater states as four basis
states of an effective Hamiltonian, the $4\times4$ secular matrix. The
components of this effective Hamiltonian are the expectation values of the
original Hamiltonian (\ref{H_K}) between the Slater states.

The two FAIR states $a_{0}^{\dag}=%
{\textstyle\sum_{\nu}}
\alpha_{0}^{\nu}c_{\nu}^{\dag}$ and $b_{0}^{\dag}=%
{\textstyle\sum_{\nu}}
\beta_{0}^{\nu}c_{\nu}^{\dag}$ determine uniquely the full basis $\left\{
a_{i}^{\dag}\right\}  $ and $\left\{  b_{i}^{\dag}\right\}  $. The two FAIR
states $a_{0}^{\dag}$ and $b_{0}^{\dag}$ are numerically optimized and
determine the secular matrix and the ground state $\Psi_{K}$.

Spin-flip processes are only included in the matrix elements between the FAIR
states $a_{0,\sigma}^{\dag},b_{0,\sigma}^{\dag}$ and the $d_{-\sigma}^{\dag}$
states. In the following I will refer to this effective Hamiltonian - secular
matrix as the "\textbf{secular Hamiltonian}".

If one applies a magnetic field $B$ in z-direction then the energy of
electrons with their moment pointing upwards is reduced by $E_{B}=\mu_{B}B$
while those with their moment pointing down is increased by $E_{B}$. Since
within this paper the orientation of the magnetic moment of the electrons is
of primary interest I use the symbol $\uparrow$ for magnetic moment up or in
short MM-up and $\downarrow$ for MM-down.

Applying a magnetic field corresponds to an external Hamiltonian%
\[
H^{ex}=-E_{B}%
{\textstyle\sum_{\nu}}
\left[  \left(  \widehat{n}_{\nu\uparrow}-\widehat{n}_{\nu\downarrow}\right)
+\left(  \widehat{n}_{d\uparrow}-\widehat{n}_{d\downarrow}\right)  \right]
\]
where $\widehat{n}_{\nu\sigma}$ is the number operator for the conduction
electron (Wilson) states. The interaction of the magnetic field with the
conduction electrons yields essentially the Pauli susceptibility in which we
are not interested at this point. Since the magnetic field is introduced in
linear response its effect on the conduction electrons can be ignored.
However, for finite magnetic fields this is not correct as will be discussed
in the appendix together with a sketch of the four Slater states in a magnetic field.

Since the effect of the magnetic field on the conduction electrons can be
discarded the effective perturbation Hamiltonian of the magnetic field has the
form%
\[
H^{ex}=-E_{B}\left(  \widehat{n}_{d\uparrow}-\widehat{n}_{d\downarrow}\right)
=E_{B}\left(  \widehat{n}_{d\downarrow}-\widehat{n}_{d\uparrow}\right)
\]

\section{Linear Response to a Magnetic Field}

The FAIR ground state (\ref{Psi_K'}) of the Kondo Hamiltonian is an exact
eigenstate of the (approximate) secular Hamiltonian (which is given in the
appendix for $J=0.1$). The ground state has the coefficients $\left(
B,B,C,C\right)  $ (with $B=0.70313,$ $C=0.0626$ for $J=0.1$).

If we apply a magnetic field $B$ which corresponds to a magnetic energy
$E_{B}$ then we obtain a polarization in our system. The perturbation
Hamiltonian in the Heisenberg picture is
\[
H_{H}^{ex}\left(  t^{\prime}\right)  =E_{B}e^{iHt^{\prime}/\hbar}\left(
\widehat{n}_{d\downarrow}-\widehat{n}_{d\uparrow}\right)  e^{-iHt^{\prime
}/\hbar}%
\]
If one switches on the field at the time $t_{0}$ then this alters the wave
function by adding a perturbation state%
\[
\left\vert \delta\Psi_{K}\right\rangle =-\frac{i}{\hbar}e^{-iHt/\hbar}%
\int_{t_{0}}^{t}H_{H}^{ex}\left(  t^{\prime}\right)  dt^{\prime}\left\vert
\Psi_{K}\left(  0\right)  \right\rangle
\]
where $\left\vert \Psi_{K}\left(  0\right)  \right\rangle $ is the ground
state in zero magnetic field in the Schroedinger picture.

Then we obtain%
\[
\left\vert \delta\Psi_{K}\right\rangle =-\frac{i}{\hbar}e^{-iHt/\hbar}%
E_{B}\int_{t_{0}}^{t}dt^{\prime}e^{iHt^{\prime}/\hbar}\left[  \left(
\widehat{n}_{d\downarrow}-\widehat{n}_{d\uparrow}\right)  e^{-iHt^{\prime
}/\hbar}\left(
\begin{array}
[c]{c}%
\left(  Ba_{0\uparrow}^{\dag}d_{\downarrow}^{\dag}+Cd_{\uparrow}^{\dag
}b_{0\downarrow}^{\dag}\right)  \left\vert \mathbf{0}_{a\uparrow}%
\mathbf{0}_{b\downarrow}\right\rangle \\
+\left(  Cb_{0\uparrow}^{\dag}d_{\downarrow}^{\dag}+Bd_{\uparrow}^{\dag
}a_{0\downarrow}^{\dag}\right)  \left\vert \mathbf{0}_{b\uparrow}%
\mathbf{0}_{a\downarrow}\right\rangle
\end{array}
\right)  \right]
\]
The term in the square brackets yields
\[
\left[  \left(  Ba_{0\uparrow}^{\dag}d_{\downarrow}^{\dag}-Cd_{\uparrow}%
^{\dag}b_{0\downarrow}^{\dag}\right)  \left\vert \mathbf{0}_{a\uparrow
}\mathbf{0}_{b\downarrow}\right\rangle +\left(  Cb_{0\uparrow}^{\dag
}d_{\downarrow}^{\dag}-Bd_{\uparrow}^{\dag}a_{0\downarrow}^{\dag}\right)
\left\vert \mathbf{0}_{b\uparrow}\mathbf{0}_{a\downarrow}\right\rangle
\right]  e^{-iE_{0}t^{\prime}/\hbar}%
\]
This state is perpendicular to $\left\vert \Psi_{K}\right\rangle $ and can be
expanded in terms of the other three solutions of the $4\times4$ secular
matrix
\[
=%
{\textstyle\sum_{i=\lambda}^{3}}
\alpha_{\lambda}\left\vert \psi_{\lambda}\right\rangle e^{-iE_{0}t^{\prime
}/\hbar}%
\]
where $\left\vert \psi_{\lambda}\right\rangle $ oscillates with the frequency
$E_{\lambda}/\hbar$. This yields for
\[
\int_{t_{0}}^{t}dt^{\prime}e^{iHt^{\prime}/\hbar}\left[  ...\right]
e^{-iE_{0}t^{\prime}/\hbar}=%
{\textstyle\sum_{\lambda=1}^{3}}
\frac{i\hbar}{\left(  E_{0}-E_{\lambda}+i\eta\hbar\right)  }\alpha_{\lambda
}\left\vert \psi_{\lambda}\right\rangle e^{-i\left(  E_{0}-E_{\lambda}\right)
t/\hbar}%
\]
Here a factor $e^{\eta t^{\prime}}$ with $\eta=0^{+}$ is inserted for
convergence reasons. Then the lower integration limit yields zero for
$t_{0}->-\infty.$ (Afterwards $\eta$ is discarded).

Then one obtains for the linear response wave function%
\[
\left\vert \delta\Psi_{K}\left(  t\right)  \right\rangle =-E_{B}%
{\textstyle\sum_{\lambda=1}^{3}}
\frac{1}{E_{\lambda,0}}\alpha_{\lambda}\left\vert \psi_{\lambda}\right\rangle
e^{-iE_{0}t/\hbar}%
\]
where $E_{\lambda,0}=-\left(  E_{0}-E_{\lambda}\right)  >0$

\subsection{Magnetic moment}

Next we calculate the magnetic moment (in units of $\mu_{B}$). The operator in
the Schroedinger picture is $\left(  \widehat{n}_{d\uparrow}-\widehat
{n}_{d\downarrow}\right)  $ The expectation value of the moment $\left\langle
\mu\right\rangle $ is%
\[
\left\langle \mu\right\rangle =2\operatorname{Re}\left\langle \delta\Psi
_{K}\left(  t\right)  \left\vert \left(  \widehat{n}_{d\uparrow}-\widehat
{n}_{d\downarrow}\right)  \right\vert \Psi_{K}\left(  t\right)  \right\rangle
\]
We calculate
\[
\left\langle \delta\Psi_{K}\left(  t\right)  \left\vert \left(  \widehat
{n}_{d\uparrow}-\widehat{n}_{d\downarrow}\right)  \right\vert \Psi_{K}\left(
t\right)  \right\rangle =
\]%
\[
=\left\langle
{\textstyle\sum_{\lambda=1}^{3}}
\frac{-E_{B}}{E_{\lambda,0}}\alpha_{\lambda}\psi_{\lambda}e^{-iE_{0}t/\hbar
}\left\vert \left(  \widehat{n}_{d\uparrow}-\widehat{n}_{d\downarrow}\right)
\right\vert \Psi_{K}\left(  0\right)  e^{-iE_{0}t/\hbar}\right\rangle
\]%
\[
=\left\langle
{\textstyle\sum_{\lambda=1}^{3}}
\frac{E_{B}}{E_{\lambda,0}}\alpha_{\lambda}\psi_{\lambda}|\left[
{\textstyle\sum_{\lambda=1}^{3}}
\alpha_{\lambda}\left\vert \psi_{\lambda}\right\rangle \right]  \right\rangle
\]
While the four Slater states in equ. (\ref{Psi_K}) are not orthogonal with
respect to each other, the eigenstates $\left\vert \psi_{\lambda}\right\rangle
$ of the secular Hamiltonian are. Therefore one obtains for the magnetic
moment%
\begin{equation}
\left\langle \mu\right\rangle =2E_{B}%
{\textstyle\sum_{\lambda=1}^{3}}
\frac{1}{E_{\lambda,0}}\left\vert \alpha_{\lambda}\right\vert ^{2} \label{mu}%
\end{equation}

For the example with the exchange interaction strength $J=0.1$ I have
collected the secular Hamiltonian, its eigenvectors and its eigenvalues in the
appendix. One obtains for the linear response wave function $\left\vert
\delta\Psi_{K}\left(  0\right)  \right\rangle =0.984\left\vert \psi
_{1}\right\rangle +0.158\,\left\vert \psi_{3}\right\rangle $. Because of the
energy denominators ($E_{1}-E_{0}=2.38\times10^{-4}$, $E_{3}-E_{0}=0.522\,$)
the contribution of $\left\vert \psi_{3}\right\rangle $ to the magnetic moment
is almost $10^{-5}$ smaller and one obtains for $\left\langle \mu\right\rangle
$ in good approximation%
\[
\left\langle \mu\right\rangle =2E_{B}\frac{1}{E_{1,0}}\left\vert \alpha
_{1}\right\vert ^{2}=\frac{\left(  0.984\right)  ^{2}}{2.38\times10^{-4}%
}2E_{B}=8140\ast E_{B}%
\]

\subsection{Kondo temperature}

In numerical normalization group (NRG) calculations one uses the
susceptibility to define the Kondo temperature as $k_{B}T_{K}=1/\left(
4\chi\right)  $. The susceptibility is, in linear response, $\chi=\left\langle
\mu\right\rangle /\left(  E_{B}\right)  $. If one applies this method to the
example with $J=0.1$ one obtains%
\[
k_{B}T_{K}=\frac{E_{B}}{4\left\langle \mu\right\rangle }=3.\,\allowbreak
07\times10^{-5}%
\]

In the past we have used the energy difference between the singlet ground
state and the relaxed triplet state for the definition of the Kondo energy.
For $J=0.1$ this yields $k_{B}T_{K}^{\ast}=2.36\times10^{-5}$. These two
values are surprisingly close together if one considers that there is some
arbitrariness in the definition of the Kondo temperature and that almost every
theoretical approach to the Kondo problem has a different definition.

\section{Net Spin Polarization}

The ground state of the Kondo impurity does not possess any spin- or magnetic
moment polarization because of its symmetry between MM-up and down states.
Therefore one has to separate the ground state mathematically into its two
magnetic components to answer the question about the existence of a Kondo
cloud. The author showed that there is a Kondo cloud in the magnetic
components of the Kondo ground state. But this polarization cloud can not be
observed in the ground state. To measure the Kondo cloud one has to disturb
the ground state sufficiently so that the symmetry between MM-up and down is
removed but not too much so that the Kondo state is not destroyed. A magnetic
field of the right strength might do the job.

In linear response the interference between the ground state $\Psi_{K}$ and
the linear response state $\delta\Psi_{LS}$ gives the change of the electron
density and electron polarization due to a small magnetic field. The
calculation of this interference is quite similar to the author's calculation
of the Kondo cloud in the previous paper \cite{B177}.

When one uses the Slater state basis $\left(  \Psi_{ad},\Psi_{da},\Psi
_{db},\Psi_{bd}\right)  $ the ground state $\Psi_{K}=\psi_{0}$ has the
components $\left(  B,B,C,C\right)  $ and the linear response state
$\delta\Psi=\alpha_{1}\left(  2E_{B}/\Delta E_{01}\right)  \psi_{1}$ has the
components
\[
\alpha_{1}\frac{2E_{B}}{\Delta E_{01}}\left(  B^{\prime},-B^{\prime
},-C^{\prime},C^{\prime}\right)
\]
where $\left(  B^{\prime},-B^{\prime},-C^{\prime},C^{\prime}\right)  $ is the
eigenvector of the first excited state $\left\vert \psi_{1}\right\rangle $ and
for $J=0.1$ the components have the values $B=0.70313,$ $C=0.0626$ and
$B^{\prime}=0.7055$, $C\prime=0.0635$ (see appendix).

If we denote the electron density between two Slater states $\left\vert
\Psi_{\alpha}\right\rangle $ and $\left\vert \Psi_{\beta}\right\rangle $ as
$\left\langle \Psi_{\alpha}\left\vert \widehat{\rho}\left(  r\right)
\right\vert \Psi_{\beta}\right\rangle =\rho_{\alpha,\beta}\left(  r\right)  $
then the total change of density due to the magnetic field is%
\begin{equation}
\delta\rho\left(  r\right)  =\alpha_{1}\frac{4E_{B}}{\Delta E_{01}}\left(
\begin{array}
[c]{c}%
BB^{\prime}\left(  \rho_{ad,ad}-\rho_{da,da}\right)  +CC^{\prime}\left(
\rho_{db,db}-\rho_{bd,bd}\right) \\
+\left(  BC^{\prime}+CB^{\prime}\right)  \left(  \rho_{ad,db}-\rho
_{da,bd}+\rho_{da,db}-\rho_{ad,bd}\right)  \allowbreak
\end{array}
\right)  =0 \label{d_ro}%
\end{equation}

Since the state $\left\vert \Psi_{da}\right\rangle $ is just the state
$\left\vert \Psi_{a,d}\right\rangle $ with reversed spins the two states have
the same charge density and the opposite spin density or polarization. The
same applies for the pairs $\left(  \rho_{db,db};\rho_{bd,bd}\right)  ,$
$\left(  \rho_{ad,db};\rho_{da,bd}\right)  $ and $\left(  \rho_{da,db}%
;\rho_{ad,bd}\right)  $. Therefore one realizes that there is no change of the
charge density. On the other hand the polarization which is defined as
$p_{\alpha,\beta}=\rho_{\uparrow;\alpha,\beta}-\rho_{\downarrow;\alpha,\beta}$
reverses sign when the spins are reversed. Therefore one has for example
$p_{\alpha,\beta}=-p_{\beta,\alpha}$. This yields a total polarization of
\begin{equation}
p\left(  r\right)  =\alpha_{1}\frac{8E_{B}}{\Delta E_{01}}\left(  BB^{\prime
}p_{ad,ad}+CC^{\prime}p_{db,db}+\left(  BC^{\prime}+CB^{\prime}\right)
\left(  p_{ad,db}+p_{da,db}\right)  \allowbreak\right)  \label{p_x}%
\end{equation}
If one divides by $\left\langle \mu\right\rangle =\alpha_{1}^{2}2E_{B}/\Delta
E_{10}$ this yields for our example $\left(  J=0.1\right)  $%
\begin{equation}
\frac{p\left(  r\right)  }{4\left\langle \mu\right\rangle }=\left[
0.504\,\,p_{ad,ad}+4.\,\allowbreak1\times10^{-3}p_{db,db}+9.\,\allowbreak
2\times10^{-2}\left(  p_{ad,db}+p_{da,db}\right)  \allowbreak\right]
\label{p_x'}%
\end{equation}
where $BB^{\prime}/\alpha_{1}=0.504$.

The calculation of the charge density of the Slater states is described in
detail in ref. \ref{B177}. A summary is given in the appendix. The calculation
can be performed for the one, two or three dimensional case. Since the Kondo
impurity couples only to a single angular momentum of the conduction electrons
(for example $l=0$) it represents essentially a one-dimensional problem. If
one defines the charge density as a one-dimensional density $\delta q/\delta
r$ (for example in three dimension the charge $\delta q$ between two spheres
of radii $r$ and $r+\delta r$) then the calculation and the results are
essentially identical in all dimensions. (The main difference is that the
oscillatory part of the charge shows a dimensional phase shift of $D\frac{\pi
}{2},$ see equ. (\ref{fr_osc})).

The four Slater states of the ground state and the linear response state are
composed of Wilson states $\psi_{\nu}\left(  \xi\right)  $. Here $\xi$ is the
the distance from the impurity in units of half the Fermi length $\lambda
_{F}/2$ so that $\xi=r/\left(  \lambda_{F}/2\right)  $. The explicit form of
$\psi_{\nu}\left(  \xi\right)  $ (using the logarithmic energy scale with
$\Lambda=2$, see appendix) for $\nu<\left(  N/2-1\right)  $ is%
\[
\psi_{\nu}\left(  \xi\right)  =2\sqrt{2^{\nu+2}}\frac{\sin\left(  \pi\xi
\frac{1}{2^{\nu+2}}\right)  }{\pi\xi}\cos\left(  \pi\xi\left(  1-\frac
{3}{2^{\nu+2}}\right)  \right)
\]
This state has a fast-oscillating component $\cos\left(  \pi\xi\left(
1-\frac{3}{2^{\nu+2}}\right)  \right)  $ (in space) and a slowly varying
component $\sin\left(  \pi\xi\frac{1}{2^{\nu+2}}\right)  $. The first
component yields the Friedel and RKKY oscillations while the second part
yields the background electron density which is given by
\[
\rho_{\nu}^{0}\left(  \xi\right)  =\left\vert \psi_{\nu}\left(  \xi\right)
\right\vert ^{2}=2^{\nu+3}\frac{\sin^{2}\left(  \pi\xi\frac{1}{2^{\nu+2}%
}\right)  }{\left(  \pi\xi\right)  ^{2}}%
\]
The appendix gives a more detailed discussion of the calculation of density
and polarization.

I will divide the calculation into two parts, (i) averaging over the
fast-oscillating contribution to calculate the overall spin polarization and
(ii) focus on the fast-oscillating part which will yield oscillations of the
spin polarization.

\subsection{Background polarization}

Since $B$ and $B^{\prime}$ are much larger than $C$ and $C^{\prime}$ the
dominant contribution comes from the first term in equ. (\ref{p_x}). The
prefactor of the second term is smaller by a factor 100. The prefactors of
$p_{db}$ and $p_{bd}$ are smaller by a factor 10. However, their contribution
is in addition reduced because the bases for MM-up (and MM-down) are different
in the two states and therefore the interference terms are essentially reduced
by the square of the multi-scalar product $\left\langle
{\textstyle\prod\limits_{j=1}^{n-1}}
a_{i}^{\dag}\Phi_{0}|%
{\textstyle\prod\limits_{j=1}^{n-1}}
b_{i}^{\dag}\Phi_{0}\right\rangle $. This multi-scalar product is about
$0.14,$ which reduces the contribution of $p_{db}$ and $p_{bd}$ by an
additional factor of about $1/50$. Therefore these terms will be neglected in
the calculation. For the discussed example $\left(  J=0.1\right)  $ the
polarization is given to good accuracy by
\[
p\left(  \xi\right)  \thickapprox2\left\langle \mu\right\rangle p_{ad,ad}%
=-\left\langle \mu\right\rangle \left(  p_{da,da}-p_{ad,ad}\right)
\]

In Fig.1 the integrated polarization is plotted in units $2\left\langle
\mu\right\rangle $ as a function of the $\log_{2}\xi$ where $\xi
=2r/\lambda_{F}$ is the distance from the impurity in units of half the Fermi
wave length.%
\begin{align*}
&
{\includegraphics[
height=3.2578in,
width=3.8597in
]%
{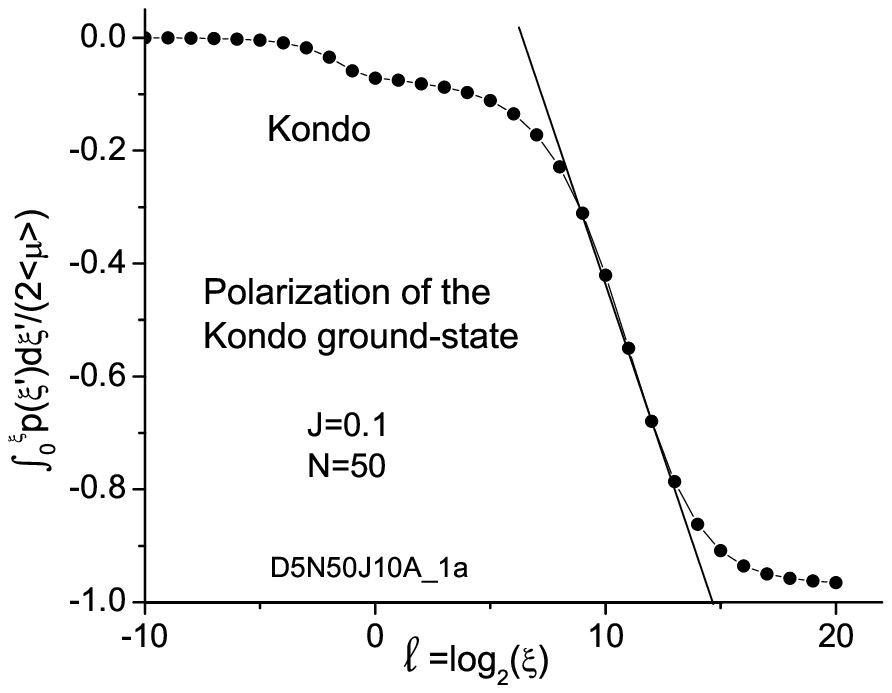}%
}%
\\
&
\begin{tabular}
[c]{l}%
Fig.1: The integrated polarization of the ground state in a magnetic\\
field in units of $2\left\langle \mu\right\rangle $ where $\mu$ is the
magnetization in units $\mu_{B}$.
\end{tabular}
\end{align*}

\subsection{Polarization oscillations}

In the vicinity of a Kondo impurity one obtains Friedel oscillations which
have the magnitude (in reduced units)
\begin{equation}
\rho_{Fr}\left(  \xi\right)  -\rho_{0}=-\frac{C_{D}}{\xi^{D}}A_{\rho}\left(
\frac{\xi}{\xi_{K}}\right)  \cos\left(  2\pi\xi-D\frac{\pi}{2}\right)
\label{fr_osc}%
\end{equation}
where $D$ is the dimension of the system, the coefficients $C_{D}$ have the
values $C_{1}=1/\left(  2\pi\right)  $, $C_{2}=1/\left(  2\pi^{2}\right)  $
and $C_{3}=1/\left(  4\pi^{2}\right)  $ in one, two and three dimensions
\cite{A83}. $r_{K}=\hbar v_{F}/k_{B}T_{K}$ is the Kondo length. The period in
$\xi$ is equal to $1$. The function $A_{\rho}\left(  \xi/\xi_{K}\right)  $ is
a universal function which approaches the values $0$ for $\xi/\xi_{K}<<1$ and
$2$ for $\xi/\xi_{K}>>1$. (I skipped the phase shift $\delta_{P}$ due to
potential scattering). I will use the magnitude of the Friedel oscillations as
a scale for the polarization oscillations in a magnetic field. Therefore I
define an amplitude $A_{p}\left(  \xi/\xi_{K}\right)  $ so that the calculated
polarization in one dimension is given by
\begin{equation}
p\left(  \xi\right)  =-\frac{1}{2\pi\xi}A_{p}\left(  \frac{\xi}{\xi_{K}%
}\right)  \cos\left(  \pi\xi-\frac{\pi}{2}-\delta_{p}\right)  \label{RKKY}%
\end{equation}

The calculated polarization oscillations are proportional to $-\cos\left(
2\pi\xi\right)  $. They are shown in Fig.4a-e for different distances from the
impurity. Each time two periods are plotted. The average distances in Fig.2a-e
vary between $\xi=7\ $and $\xi=32767$. The minima lie at integer values of
$\xi$ and the maxima at half integer ($\xi=2r/\lambda_{F}$).%

\begin{align*}
&
\begin{array}
[c]{cc}%
{\includegraphics[
height=2.4309in,
width=2.8891in
]%
{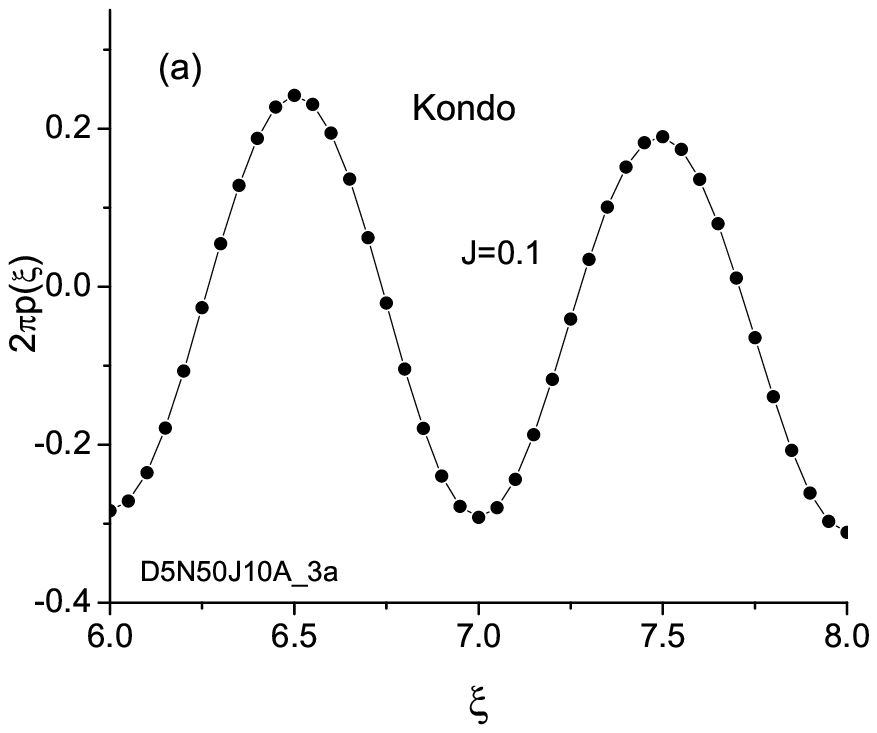}%
}%
&
{\includegraphics[
height=2.4168in,
width=2.8925in
]%
{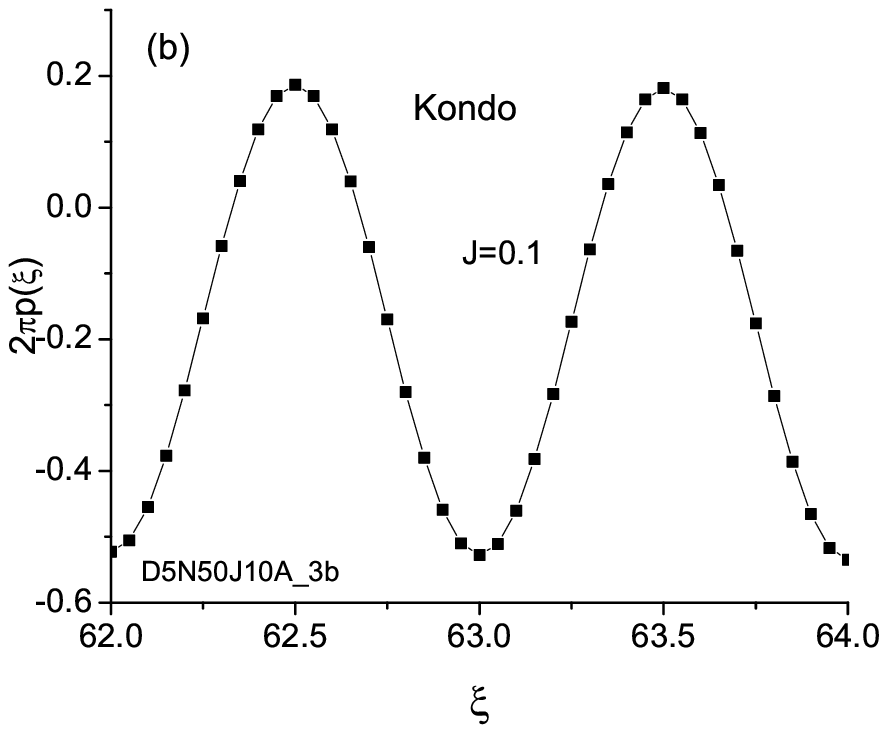}%
}%
\\%
{\includegraphics[
height=2.4026in,
width=2.8759in
]%
{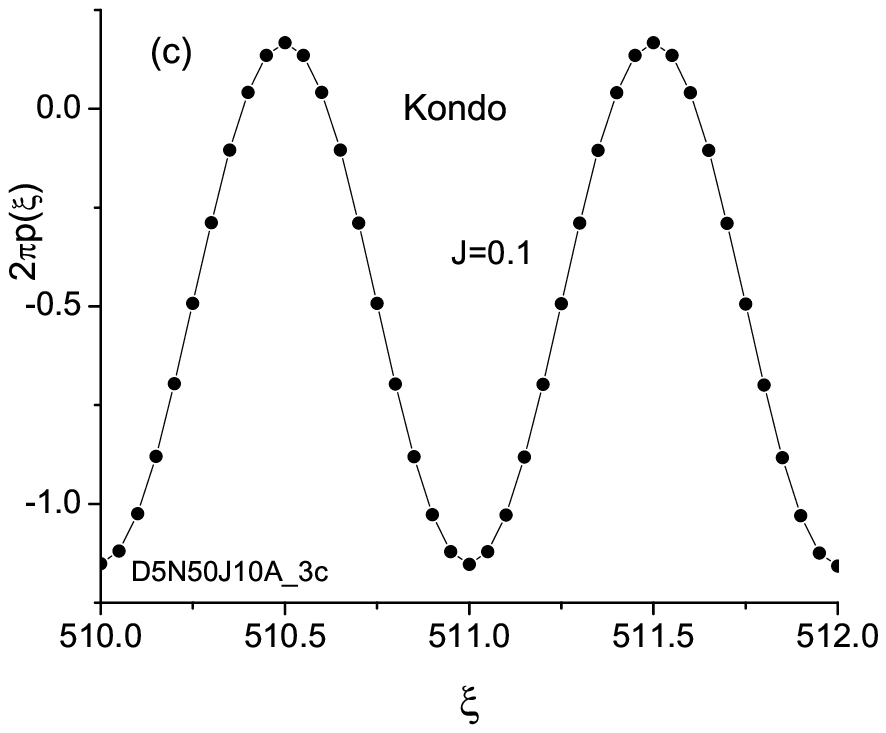}%
}%
&
{\includegraphics[
height=2.4475in,
width=2.9597in
]%
{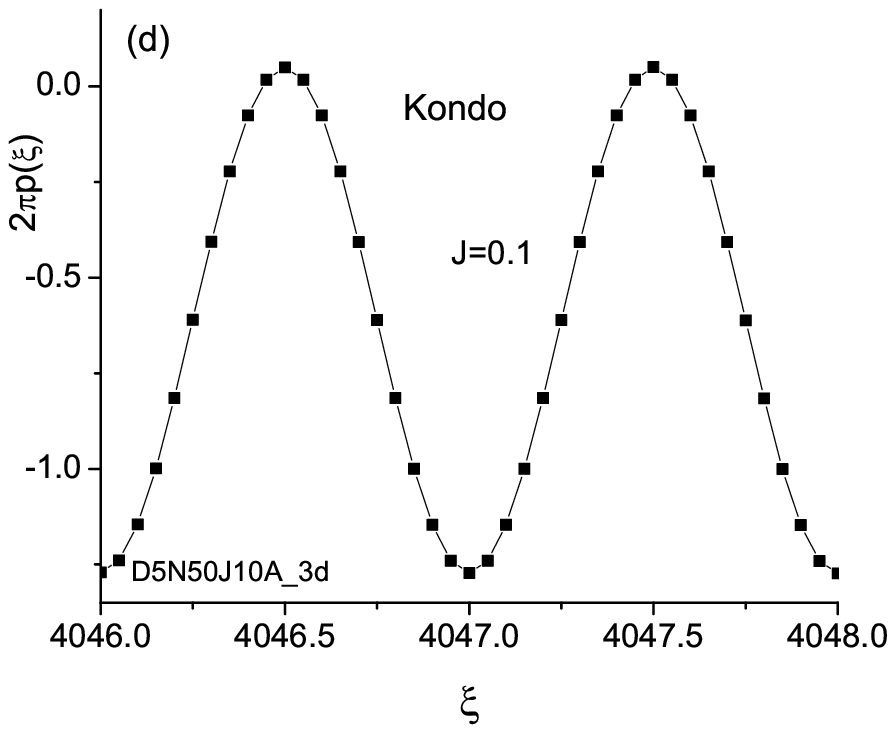}%
}%
\\%
{\includegraphics[
height=2.455in,
width=2.9489in
]%
{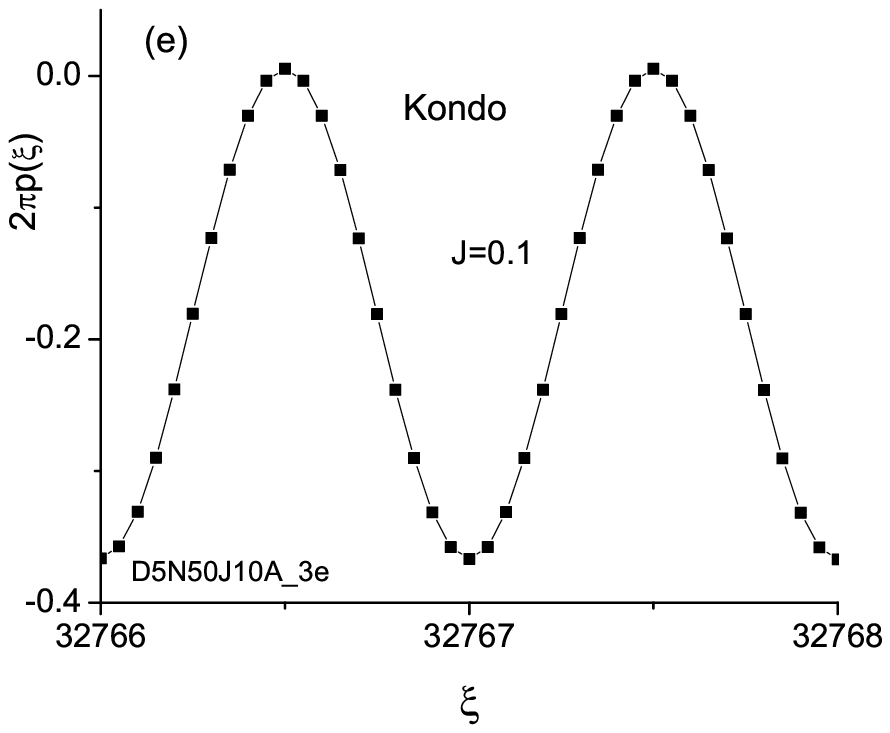}%
}%
&
\end{array}
\\
&
\begin{tabular}
[c]{l}%
Fig.2: Oscillations of $2\pi p\left(  \xi\right)  $ ($p\left(  \xi\right)  $
is the magnetic polarization) as a function \\
of $\xi=2r/\lambda_{F}$. Two two periods of the oscillations are shown for
different \\
distances from the impurity. The average distances are from (a) to (e): \\
$7,$ $63,$ $511,$ $4047$ and $32767$.
\end{tabular}
\end{align*}

In Fig.3 the envelope of the oscillations of $2\pi\xi\ast p\left(  \xi\right)
$, i.e. the maxima and minima are plotted as triangles. The full points give
the average of $2\pi\xi\ast p\left(  \xi\right)  $ as a function $l=\log
_{2}\left(  \xi\right)  $. The amplitude is given by half the difference
between the maximum and minimum curve. The amplitude does not show the
expected long-range behavior of $p\left(  \xi\right)  \backsimeq1/\left(
2\pi\xi\right)  $ (which should be a constant in Fig.2). For small distances
the average is close to zero and for distances of the order of the Kondo
length the amplitude is roughly equal to the average. For distances larger
than the Kondo length the oscillations fade away. As discussed below they
don't extend far beyond the Kondo length.%
\begin{align*}
&
{\includegraphics[
height=3.3167in,
width=3.9493in
]%
{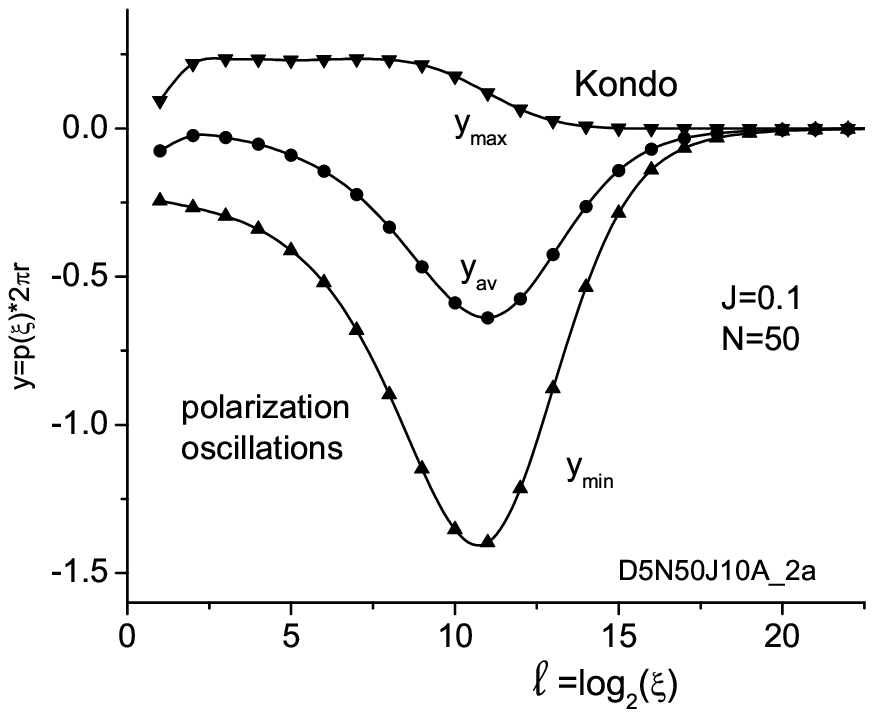}%
}%
\\
&
\begin{tabular}
[c]{l}%
Fig.3: The up and down triangles show the envelope of the oscillations of \\
$2\pi\xi\ast p\left(  \xi\right)  $ where $p\left(  \xi\right)  $ is the
magnetic polarization of the conduction electrons. \\
In the normalized coordinate $\xi$ (distance from the impurity) the
oscillation \\
period is 1. Since the abscissa is given in $l=\log_{2}\xi$ there are $2^{15}$
oscillations \\
in the range $2\leq l\leq15$. The amplitude of $2\pi\xi\ast p\left(
\xi\right)  $ does not approach a\\
constant at large distances.
\end{tabular}
\end{align*}%
\[
\]

\section{Discussion and Conclusion}

A magnetic field alters the ground state of the Kondo impurity and yields in
linear response
\[
\Psi_{K}=B\left(  1-\varepsilon\right)  \Psi_{a,d}+B\left(  1+\varepsilon
\right)  \Psi_{da}+C\left(  1+\varepsilon^{\prime}\right)  \Psi_{db}+C\left(
1-\varepsilon^{\prime}\right)  \Psi_{bd}%
\]
with
\[%
\begin{array}
[c]{ccc}%
\varepsilon=\alpha_{1}\frac{2E_{B}}{\Delta E_{01}}\frac{B^{\prime}}%
{B}=1.\,\allowbreak013\,\,\left\langle \mu\right\rangle  &  & \varepsilon
^{\prime}=\alpha_{1}\frac{2E_{B}}{\Delta E_{01}}\frac{C^{\prime}}%
{C}\,=1.\,\allowbreak031\left\langle \mu\right\rangle
\end{array}
\]
Therefore one has $\varepsilon\thickapprox\varepsilon^{\prime}\thickapprox
\left\langle \mu\right\rangle $ and the components with the d magnetic moment
parallel to the magnetic field are relatively increased in their amplitude by
the value of the magnetization $\left\langle \mu\right\rangle $. Therefore the
part of the Kondo cloud which is due to $\Psi_{da}$ is now only partically
compensated by $\Psi_{a,d}.$ The same applies to $\Psi_{db}$ and $\Psi_{bd}$.
Therefore it is not surprizing that the net polarization has the same shape as
the hidden Kondo cloud which the author calculated recently. One even expects
the scaling factor of about $2\,\left\langle \mu\right\rangle $. The
polarization cloud extends to a distance of $2^{14.9}$ which \allowbreak
corresponds to $1.5\times10^{4}\lambda_{F}$ being of the order of the Kondo
length. Using the Kondo energy $k_{B}T_{K}\thickapprox3.1\times10^{-5}$ one
obtains for the Kondo length $r_{K}=\hbar v_{F}/k_{B}T_{K}$ $=\lambda
_{F}/\left(  2\pi k_{B}T_{K}\right)  $ $\thickapprox5.1\times10^{3}\lambda
_{F}$ where Wilson's linear dispersion relation between energy and wave vector
is used. \ 

On the other hand the position dependence of the polarization oscillation is
at first rather surprizing. The amplitude of $\left(  C_{D}/\xi^{D}\right)
\cos\left(  \pi\xi\right)  $ has a maximum at roughly the position
$\xi\thickapprox2^{10.5}=\allowbreak1.45\times10^{3}\ $which is smaller than
the Kondo length but corresponds roughly to the distance at which the Friedel
oscillation began to develop. Since the Friedel oscillations are supressed at
short distances it is not surprizing that the polarization osccillations are
suppressed within this range. But why do they disapper at larger distances.
Here one can consider two physical scenarios:

\begin{itemize}
\item The magnetic field polarizes the Kondo impurity and generates a magnetic
moment at the impurity. This magnetic moment scatters MM-up and down electrons
differently and as a result one obtains RKKY oscillations which should depend
as $\xi^{-D}$ on the distance from the impurity.

\item The Kondo impurity modifies the electronic structure of the host within
the distance of the Kondo impurity. The polarization oscillations are
restricted to the region into which the Kondo cloud of the impurity extends.
\end{itemize}

It appears that the polarization oscillations (for small magnetic fields) are
not due to the scattering of conduction electrons by the impurity. Instead
they show the structure of the Kondo cloud and therefore they are restricted
to distances of the order of the Kondo radius.

\section{Appendix}

\subsection{The FAIR Method}

The ansatz for the ground state of the Kondo Hamiltonian (equ.(\ref{Psi_K})
has been derived from the ground state of the Friedel-Anderson impurity. The
Hamiltonian of the Friedel-Anderson impurity is given
\[
H_{FA}=%
{\textstyle\sum_{\sigma}}
\left\{  \sum_{\nu=0}^{N-1}\varepsilon_{\nu}c_{\nu\sigma}^{\dag}c_{\nu\sigma
}+E_{d}d_{\sigma}^{\dag}d_{\sigma}+\sum_{\nu=0}^{N-1}V_{sd}(\nu)[d_{\sigma
}^{\dag}c_{\nu\sigma}+c_{\nu\sigma}^{\dag}d_{\sigma}]\right\}  +U\widehat
{n}_{d\uparrow}\widehat{n}_{d\downarrow}%
\]
In mean-field approximation the Hamiltonian one replaces $n_{d+}n_{d-}$ =%
$>$%
$n_{d+}\left\langle n_{d-}\right\rangle $ $+\left\langle n_{d+}\right\rangle
n_{d-}$ $-\left\langle n_{d+}\right\rangle \left\langle n_{d-}\right\rangle $.
After adjusting $\left\langle n_{d+}\right\rangle $ and $\left\langle
n_{d-}\right\rangle $ self-consistently one obtains two Friedel resonance
Hamiltonians with a spin-dependent energy of the $d_{\sigma}$-state:
$E_{d,\sigma}$ $=E_{d}+U\left\langle n_{d,-\sigma}\right\rangle $. The
mean-field wave function is a product of two Friedel ground states for spin up
and down $\Psi_{mf}=\Psi_{F\uparrow}\Psi_{F\downarrow}$ .

As discussed in section II the ground state of the Friedel Hamiltonian can be
expressed as a function of two Slater states, one consisting of only
conduction electrons and the other containing one d-electron.
\[
\Psi_{F}=\left(  Aa_{0}^{\dag}+Bd^{\dag}\right)
{\textstyle\prod\limits_{i=1}^{n-1}}
a_{i}^{\dag}\Phi_{0}%
\]
The state $a_{0}^{\dag}$ is called the FAIR state.

The product of the two Friedel ground states in the mean field approximation
can be expanded and yields a "magnetic state"%

\begin{equation}
\Psi_{MS}=\left[  Aa_{0\uparrow}^{\dag}b_{0\downarrow}^{\dag}+Ba_{0\uparrow
}^{\dag}d_{\downarrow}^{\dag}+Cd_{\uparrow}^{\dag}b_{0\downarrow}^{\dag
}+Dd_{\uparrow}^{\dag}d_{\downarrow}^{\dag}\right]  \left\vert \mathbf{0}%
_{a\uparrow}\mathbf{0}_{b\downarrow}\right\rangle \label{Psi_MS}%
\end{equation}
where $\left\{  a_{i}^{\dag}\right\}  $ and $\left\{  b_{i}^{\dag}\right\}  $
are two (different) bases of the $N$-dimensional Hilbert space and
$a_{0}^{\dag},b_{0}^{\dag}$ are two different FAIR states.%
\[
\left\vert \mathbf{0}_{a\uparrow}\mathbf{0}_{b\downarrow}\right\rangle
=\prod_{j=1}^{n-1}a_{j\uparrow}^{\dag}\prod_{j=1}^{n-1}b_{j\downarrow}^{\dag
}\left\vert \Phi_{0}\right\rangle
\]
The state (\ref{Psi_MS}) opens a wide playing field for optimization: (i) The
FAIR states $a_{0}^{\dag}$ and $b_{0}^{\dag}$ can be individually optimized to
minimize the energy expectation value (ground-state energy). Each FAIR state
defines uniquely a whole basis $\left\{  a_{i}^{\dag}\right\}  ,$ $\left\{
b_{i}^{\dag}\right\}  $. The coefficients $A,B,C,D$ can be optimized. This
yields a much better treatment of the correlation effects. The resulting state
is denoted as the (potentially) magnetic state $\Psi_{MS}$.

The magnetic state $\Psi_{MS}$ is used as the building block for the singlet
state. $\Psi_{MS}$ together with its counterpart\ where all spins are reversed
yield two states $\overline{\Psi}_{MS}\left(  \uparrow\downarrow\right)  $ and
$\overline{\overline{\Psi}}_{MS}\left(  \downarrow\uparrow\right)  .$ The
singlet ground state is then given by
\[
\Psi_{SS}=\overline{\Psi}_{MS}\left(  \uparrow\downarrow\right)  \pm
\overline{\overline{\Psi}}_{MS}\left(  \downarrow\uparrow\right)
\]
A new optimization of the FAIR states and the coefficients changes
$a_{0}^{\dag}$ and $b_{0}^{\dag}$ drastically increases the weight at low energies.

The Kondo Hamiltonian represents a limit of the Friedel-Anderson Hamiltonian
by suppressing zero and double occupancy of the d-states. This leads to the
approximate ground state \ref{Psi_K} in chapter II.

\subsection{Wilson's states}

Wilson considered an s-band with a constant density of states and the Fermi
energy in the center of the band. By measuring the energy from the Fermi level
and dividing all energies by the Fermi energy Wilson obtained a band ranging
from $-1$ to $+1.$ To treat the electrons close to the Fermi level at
$\zeta=0$ as accurately as possible he divided the energy interval $\left(
-1:0\right)  $ at energies of $-1/\Lambda,-1/\Lambda^{2},-1/\Lambda^{3},..$
i.e. $\zeta_{\nu}=-1/\Lambda^{\nu}.$ This yields energy cells $\mathfrak{C}%
_{\nu}$ with the range $\left\{  -1/\Lambda^{\nu}:-1/\Lambda^{\nu+1}\right\}
$ and the width $\Delta_{\nu}$ $=\zeta_{\nu+1}-\zeta_{\nu}$ $=1/\Lambda
^{\nu+1}$. During this paper generally the value $\Lambda=2$ is chosen. (In
this paper I count the energy cells and the Wilson states from $\nu=1$ to $N$.)

Wilson rearranged the quasi-continuous original electron states $\varphi
_{k}\left(  r\right)  $ in such a way that only one state within each cell
$\mathfrak{C}_{\nu}\ $had a finite interaction with the impurity. Assuming
that the interaction of the original electron states $\varphi_{k}\left(
r\right)  $ with the impurity is $k$-independent this interacting state in
$\mathfrak{C}_{\nu}$ had the form%
\[
\psi_{\nu}\left(  r\right)  =%
{\textstyle\sum_{\mathfrak{C}_{\nu}}}
\varphi_{k}\left(  r\right)  /\sqrt{Z_{\nu}}%
\]
where $Z_{\nu}$ is the total number of states $\varphi_{k}\left(  r\right)  $
in the cell $\mathfrak{C}_{\nu}$ ($Z_{\nu}=Z\left(  \zeta_{\nu+1}-\zeta_{\nu
}\right)  /2,$ $Z$ is the total number of states in the band). There are
$\left(  Z_{\nu}-1\right)  $ additional linear combinations of the states
$\varphi_{k}$ in the cell $\mathfrak{C}_{\nu}$ but they have zero interaction
with the impurity and were ignored by Wilson as they are within this paper.

The interaction strength of the original basis states $\varphi_{k}\left(
r\right)  $ with the d-impurity is assumed to be a constant. Then the exchange
interaction between two Wilson states $\psi_{\nu}\left(  r\right)  $ and
$\psi_{\mu}\left(  r\right)  $ is given $J_{\nu,\mu}=\sqrt{Z_{\nu}/Z}%
\sqrt{Z_{\mu}/Z}J=\sqrt{\left(  \zeta_{\nu+1}-\zeta_{\nu}\right)  /2}%
\sqrt{\left(  \zeta_{\mu+1}-\zeta_{\mu}\right)  /2}J$.

\subsection{The Slater states in a finite magnetic field}

In Fig.4 the energy bands of the conduction electrons $a_{i,\sigma}^{\dag}$
and $b_{j,\sigma}^{\dag}$ are drawn together with the FAIR states
$a_{0,\sigma}^{\dag}$ and $b_{0,\sigma}^{\dag}$ and the d states $d_{\sigma
}^{\dag}$ . We have four Slater states. The first Slater state is occupied
with $\left(  n-1\right)  $ electrons of the basis $\left\{  a_{i\uparrow
}^{\dag}\right\}  $ and $\left(  n-1\right)  $ electrons of the basis
$\left\{  b_{i\downarrow}^{\dag}\right\}  ,$the FAIR state $a_{0\uparrow
}^{\dag}$ and the d-state $d_{\downarrow}$. In the absence of a magnetic field
the second Slater state is obtained by reversing all moments. So in the second
Slater state the MM-up electrons would be filled into the $\left\{
b_{i\uparrow}^{\dag}\right\}  $ basis and the MM-down electrons into the
$\left\{  a_{i\downarrow}^{\dag}\right\}  $ basis. However, since the magnetic
field shifts MM-up and MM-down electrons in different directions, the basis
are slightly altered to $\left\{  b_{i\uparrow}^{\prime\dag}\right\}  $ and
$\left\{  a_{i\downarrow}^{\prime\dag}\right\}  $.

The energies of $a_{0}^{\dag}$ and $b_{0}^{\dag}$ are non-zero and opposite
equal. In the first two Slater states the $a_{0}^{\dag}$ state with negative
energy is occupied while in the last two Slater states the $b_{0}^{\dag}$
state with positive energy is occupied. The occupation probability of the last
two is smaller by roughly a factor 100 than for the first two.

\begin{align*}
&
{\includegraphics[
height=5.731in,
width=3.9336in
]%
{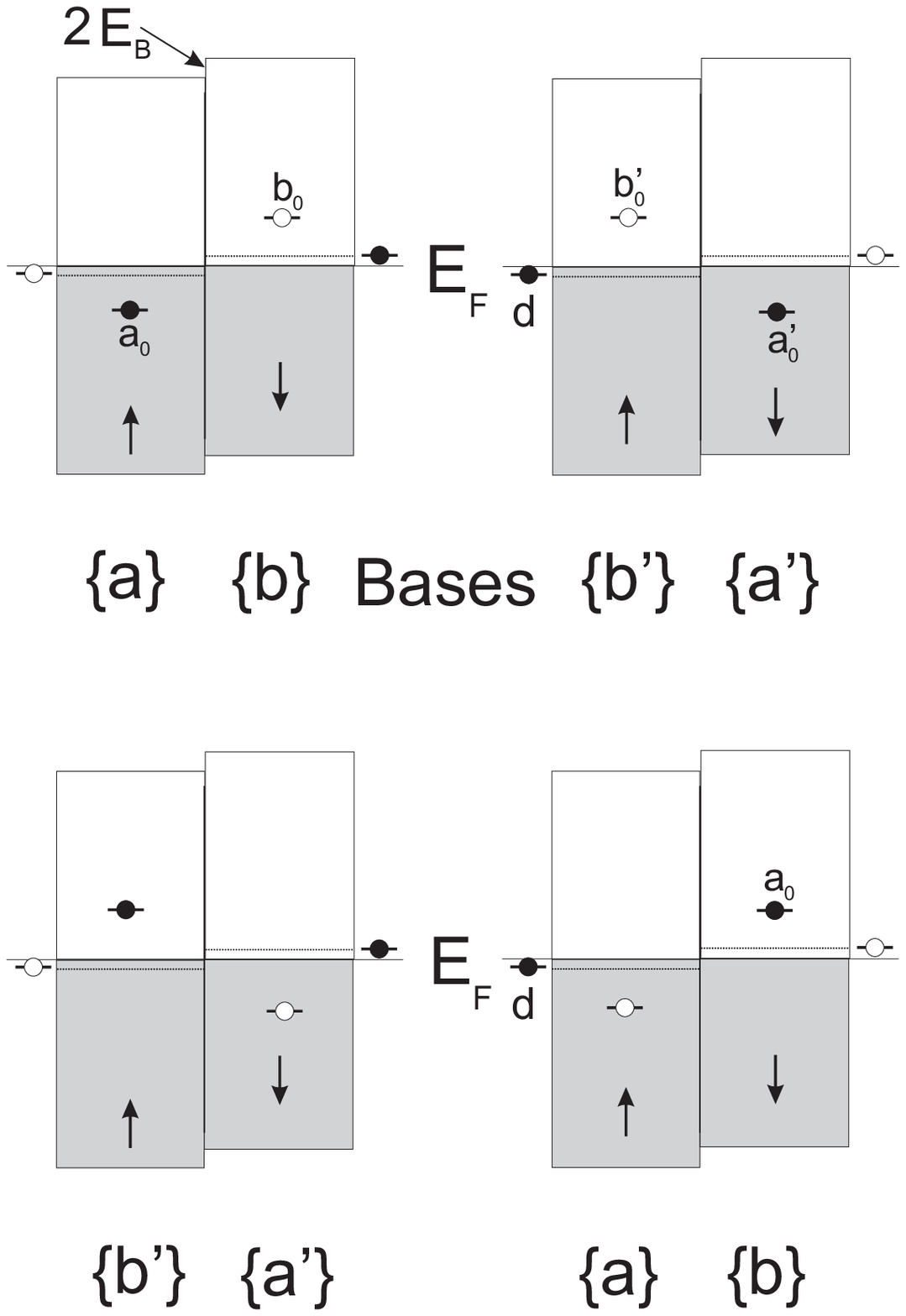}%
}%
\\
&
\begin{tabular}
[c]{l}%
Fig.4: A drawing of the four Slater states in the Kondo ground state. In
each\\
Slater state a FAIR state ($a_{0}^{\dag}$ or $b_{0}^{\dag}$) and a d-state
with opposite spins\\
(magnetic moment) \ are occupied. The energies of the FAIR states are\\
non-zero and opposite equal. The brackets below the bands show which\\
basis the Wilson states occupy. A magnetic field has shifted the moment\\
up and down bands in different directions.
\end{tabular}
\end{align*}

\[
\]
The magnetic field moves the MM-up and the MM-down bands in opposite direction
by $\pm E_{B}$. The Fermi levels of MM-up and MM-down electrons readjust by
electron spin flips to the same height. As a consequence the Fermi momenta are
slightly different for MM-up and the MM-down electrons.

In the present calculation the slight shift of the bands is neglected because
the shift is very small compared to the band energy. The Wilson states in the
shifted bands measure their energies from the new Fermi level. Then the energy
of the Wilson states is not changed except for the first and last Wilson
state. The small change for $\nu=1$ which extends from $-1\pm E_{B}$ to
$-\frac{1}{2}$ and for $\nu=N$ which extends from $+\frac{1}{2}$ to $1\pm
E_{B}$ is neglected. Since the FAIR bases are constructed from the Wilson
states their energies are, in first approximation not effected by the magnetic field.

A finite magnetic field has, however, an important effect on the matrix
elements. If one considers two Wilson states $c_{\nu\uparrow}^{\dag}$ and
$c_{\nu\downarrow}^{\dag}$ with the same energy but different MM then their
orbital wave functions are not identical. As a consequence the scalar product
between the orbital parts of $c_{\nu\uparrow}^{\dag}$ and $c_{\nu\downarrow
}^{\dag}$ is no longer equal to one but reduced to $1-2E_{B}/\left(
\zeta_{\nu+1}-\zeta_{\nu}\right)  $ (for $2E_{B}<\left(  \zeta_{\nu+1}%
-\zeta_{\nu}\right)  $). In the FAIR method such matrix elements occur only
between the FAIR states. Since the present calculation is performed in linear
response this problem does not arise. But for finite fields it will be important.

\subsection{The Secular Hamiltonian and its Eigenstates}

The FAIR solution for the Kondo ground state is given by the ansatz
(\ref{Psi_K}) which consists of four Slater states. With this state $\Psi_{K}$
one can calculate the expectation value of the Hamiltonian $H_{K}$ (equ.
(\ref{H_K})). The minimization of this expectation value with respect to the
four coefficients (which at this stage are considered as independent) yields a
$4\times4$ secular matrix. In parallel the variation and optimization of the
FAIR states $a_{0}^{\dag}$ and $b_{0}^{\dag}$ yields the optimal bases
$\left\{  a_{i}^{\dag}\right\}  $ and $\left\{  b_{i}^{\dag}\right\}  $ with
the lowest ground-state energy. (The calculation and optimization of the
energy expectation value has been described in ref. \cite{B153}).

The secular matrix possesses four eigenvectors with four energy eigenvalues.
The coefficients of the eigenvector with the lowest energy are the
coefficients of the ground state $\left\vert \psi_{0}\right\rangle =\left\vert
\Psi_{K}\right\rangle $. The other eigenvectors represent excited states
$\left\vert \psi_{1}\right\rangle $,$\left\vert \psi_{2}\right\rangle
,\left\vert \psi_{3}\right\rangle $ which are needed in the linear response
calculation. With respect to these eigenvectors and their energy the secular
matrix plays the role of a Hamiltonian, which I baptized the \textbf{secular
Hamiltonian}. For $J=0.1$ and a Wilson basis with $\Lambda=2$ and $N=50$ this
secular Hamiltonian is given below by equ. (\ref{S_H}).%

\begin{equation}
H_{\sec}=\left(
\begin{array}
[c]{cccc}%
-3.0013 & -.00019237 & -.046265 & -.057213\\
-.00019237 & -3.0013 & -.057213 & -.046265\\
-.046265 & -.057213 & -2.4887 & -.0029434\\
-.057213 & -.046265 & -.0029434 & -2.4887
\end{array}
\right)  \label{S_H}%
\end{equation}

The four eigenvectors $\allowbreak$of (\ref{S_H}) are given by the four
columns of the $4\times4$-matrix (\ref{E_V}) below.
\begin{equation}
\left(
\begin{array}
[c]{cccc}%
.70313 & .70547 & -.076084 & .049993\\
.70313 & -.70547 & -.076084 & -.049993\\
.06256 & .063539 & .70447 & -.70438\\
.06256 & -.063539 & .70447 & .70438
\end{array}
\right)  \label{E_V}%
\end{equation}
The eigenvalues are $\left(  E_{0},E_{1},E_{2},E_{3}\right)  =\left(
-3.0056,-3.0053,-2.4856,-2.4831\right)  $. The energy difference $\left(
E_{1}-E_{0}\right)  $ is $E_{1,0}=2.38\times10^{-4}$. The band energies for
the occupied states in basis $\left\{  a_{i}^{\dag}\right\}  $ is
$E_{bd}\left(  a\right)  =$ $-1.4988$ and for the basis $\left\{  b_{i}^{\dag
}\right\}  $ is $E_{bd}\left(  b\right)  =-1.22305$.

\subsection{The wave function of Wilson's states in real space}

For the discussion of the wave functions in real space we assume a linear
dispersion relation between energy and momentum, a constant density of states
and a constant amplitude at $r=0.$ Then the interaction between the original
basis states $\varphi_{k}\left(  r\right)  $ and the d-impurity will be
constant for all k-states. These assumptions are the same as in Wilson's
treatment of the Kondo impurity. We define the wave functions $\varphi
_{k}\left(  r\right)  $ in such a way that the results apply for the impurity
problem in one, two and three dimensions.

\subsubsection{One-dimensional case}

Let us start with the one-dimensional problem. Here we have the impurity at
the position zero and the conduction electrons are located in the range
between $0$ and $L.$ The wave functions $\varphi_{k}\left(  r\right)  $ have
the form $\varphi_{k}\left(  r\right)  =\sqrt{2/L}\cos\left(  kr\right)  $.
There is another set of eigenstates $\overline{\varphi_{k}\left(  r\right)
}=\sqrt{2/L}\sin\left(  kr\right)  $. These states don't interact with the
impurity at the origin. Therefore they don't have any bearing on the impurity problem.

\subsubsection{Three-dimensional case}

In three dimensions the free electron states can be expressed as $\varphi
_{k}\left(  r\right)  \varpropto Y_{l}^{m}\left(  \theta,\phi\right)
j_{l}\left(  kr\right)  $ where $Y_{l}^{m}$ is a spherical harmonics and $l,m$
are the angular momentum and magnetic quantum numbers. $j_{l}\left(
kr\right)  $ is a spherical Bessel function. Its long range behavior is given
by $\left(  1/kr\right)  \sin\left(  kr-l\pi/2\right)  $. Only the states with
the same $l$ as the impurity couple to the impurity. All the other states for
different $l$ belong to the group of inert states $\overline{\varphi
_{k}\left(  r\right)  }$.

If one calculates the density of the wave function, integrating in the
three-dimensional case over the spherical surface $4\pi r^{2}$ and averaging
over short range (Friedel) oscillations then one obtains in the

\begin{itemize}
\item one-dimensional case: $\left(  2/L\right)  \overline{\cos^{2}\left(
kr\right)  }=1/L$

\item three-dimensional case: $\rho_{k}\left(  r\right)  =\left(  2/L\right)
\overline{\sin^{2}\left(  kr-l\pi/2\right)  }=1/L$
\end{itemize}

In both cases one obtains essentially the same density. Therefore it is
sufficient to use the one-dimensional approach for calculating the density of
a Kondo cloud. It is equivalent to the 3-dimensional case integrated over the
spherical surface.

\subsubsection{The wave functions in one dimension}

While the energy is measured in units of the Fermi energy the momentum will be
measured in units of the Fermi wave number. We assume a linear dispersion
relation for $0\leq\kappa\leq2$ with
\[
\zeta=\left(  \kappa-1\right)
\]
Here $\kappa=k/k_{F}$ is dimensionless. It is useful to measure distances also
in dimensionless units. We define $\xi=\frac{1}{\pi}k_{F}r=\frac{r}%
{\lambda_{F}/2}$. Then $\xi$ gives the distance from the impurity in units of
$\lambda_{F}/2$. The (almost) continuous states $\varphi_{\kappa}$ are given
as
\[
\varphi_{\kappa}\left(  \xi\right)  =\sqrt{\frac{2}{L}}\cos\left(  \pi
\kappa\xi\right)
\]
where $L$ is the length of the one-dimensional box. The boundary condition
$\cos\left(  \pi\kappa L\right)  =0$ yields $\kappa=\left(  \lambda
+1/2\right)  /L$ (where $\lambda$ is an integer. The maximum value of
$\lambda$ is the integer of $\left(  2L\right)  ,$ since $\kappa$ is
dimensionless then $L$ is also dimensionless). Therefore we have $Z=2L$ states
in the full band of width $2$.

To obtain the Wilson state we have to sum the states $\varphi_{\kappa}\left(
\xi\right)  $ over all states within an energy cell. If the cell ranges from
$\left(  \zeta_{\nu}:\zeta_{\nu+1}\right)  $ corresponding to a $\kappa$-range
$\left(  1+\zeta_{\nu}\right)  <\kappa<\left(  1+\zeta_{\nu+1}\right)  $ then
we represent all the states in this energy interval by%
\[
\psi_{\nu}\left(  \xi\right)  =\frac{1}{\sqrt{\left(  \zeta_{\nu+1}-\zeta
_{\nu}\right)  L}}%
{\textstyle\sum_{1+\zeta_{\nu}<\kappa<1+\zeta_{\nu+1}}}
\sqrt{\frac{2}{L}}\cos\left(  \pi\kappa\xi\right)
\]
From $Z_{\nu}=L\left(  \zeta_{\nu+1}-\zeta_{\nu}\right)  $ states we have
(according to Wilson) constructed one state $\psi_{\nu}\left(  \xi\right)  $
which couples to the impurity. Similarly one can construct $\left(  Z_{\nu
}-1\right)  $ additional linear combinations of $\varphi_{\kappa}\left(
\xi\right)  $ which are orthonormal and do not couple to the impurity at the
origin. We denote these states as $\overline{\overline{\varphi_{\nu,l}\left(
\xi\right)  }}$. They are as inert to the impurity as the states
$\overline{\varphi_{\kappa}\left(  \xi\right)  }$ and will be included in the quasi-vacuum.

After integration the wave function of the state $c_{\nu}^{\dag}$ has the form
for $\nu<N/2$%
\[
\psi_{\nu}\left(  \xi\right)  =\frac{2\sqrt{2}}{\sqrt{\left(  \zeta_{\nu
+1}-\zeta_{\nu}\right)  }}\frac{\sin\left(  \frac{\pi\xi\left(  \zeta_{\nu
+1}-\zeta_{\nu}\right)  }{2}\right)  }{\pi\xi}\cos\left(  \frac{\pi\xi\left(
2+\zeta_{\nu}+\zeta_{\nu+1}\right)  }{2}\right)
\]
For the logarithmic energy scale $\left(  \Lambda=2\right)  $ this yields for
$\nu<N/2-1$ using $\zeta_{\nu}=-1/2^{\nu}$%
\begin{equation}
\psi_{\nu}\left(  \xi\right)  =2\sqrt{2^{\nu+2}}\frac{\sin\left(  \pi\xi
\frac{1}{2^{\nu+2}}\right)  }{\pi\xi}\cos\left(  \pi\xi\left(  1-\frac
{3}{2^{\nu+2}}\right)  \right)  \label{psi_nu}%
\end{equation}
Similarly one obtains for in the positive energy range
\[
\psi_{N-1-\nu}\left(  \xi\right)  =2\sqrt{2^{\nu+2}}\frac{\sin\left(  \frac
{1}{2^{\nu+2}}\pi\xi\right)  }{\pi\xi}\cos\left(  \pi\xi\left(  1+\frac
{3}{2^{\nu+2}}\right)  \right)
\]

The two wave functions $\psi_{N/2-1}$ and $\psi_{N/2}$ are special because
their $\kappa$-range is the same as their neighbors $\psi_{N/2-2}$ and
$\psi_{N/2+1}$ All four states close to the Fermi level have the same $\kappa
$-range of $2^{-N/2-1}.$ One has to pay special attention to this complication.

\subsection{Density of the Wilson states in real space}

The density of a single state $\psi_{\nu}\left(  \xi\right)  $ is given by the
square of the function $\psi_{\nu}\left(  \xi\right)  $ in equ. (\ref{psi_nu}%
).
\[
\left\vert \psi_{\nu}\left(  \xi\right)  \right\vert ^{2}=\frac{8}{\left(
\zeta_{\nu+1}-\zeta_{\nu}\right)  }\frac{\sin^{2}\left(  \frac{\pi\xi\left(
\zeta_{\nu+1}-\zeta_{\nu}\right)  }{2}\right)  }{\left(  \pi\xi\right)  ^{2}%
}\cos^{2}\left(  \frac{\pi\xi\left(  2+\zeta_{\nu}+\zeta_{\nu+1}\right)  }%
{2}\right)
\]

This density has a fast oscillating contribution which yields the Friedel
oscillations. We first average over the fast oscillation (which has a period
of the order of 1 in units of $\lambda_{F}/2)$. Then we obtain for the Wilson
states for $\nu<N/2$ where $\left(  \zeta_{\nu+1}-\zeta_{\nu}\right)
=\frac{1}{2^{\nu+1}}$%
\begin{equation}
\rho_{\nu}^{0}\left(  \xi\right)  =\left\vert \psi_{\nu}\left(  \xi\right)
\right\vert ^{2}=2^{\nu+3}\frac{\sin^{2}\left(  \pi\xi\frac{1}{2^{\nu+2}%
}\right)  }{\left(  \pi\xi\right)  ^{2}} \label{ro_nu}%
\end{equation}
In the numerical calculation we will use (most of the time) $N=50$ Wilson
states. The different wave functions $\psi_{\nu}\left(  \xi\right)  $ have
very different spatial ranges and therefore very different densities, the
lowest being of the order of $2^{-25}<3\times10^{-8}$. This means that it is
not useful to calculate the density as a function of $\xi$ because this
density varies over a range of $2^{25}$. Instead, we use the integrated
density, integrated from $0$ to $\xi$.%
\begin{align*}
q_{\nu}^{0}\left(  \xi\right)   &  =\int_{0}^{\xi}\left\vert \psi_{\nu}\left(
\xi^{\prime}\right)  \right\vert ^{2}d\xi^{\prime}=2^{\nu+3}\int_{0}^{\xi
}\frac{\sin^{2}\left(  \pi\xi^{\prime}\frac{1}{2^{\nu+2}}\right)  }{\left(
\pi\xi^{\prime}\right)  ^{2}}d\xi^{\prime}\\
&  =2\int_{0}^{\frac{\xi}{2^{\nu+2}}}\frac{\sin^{2}\left(  \pi u\right)
}{\left(  \pi u\right)  ^{2}}du
\end{align*}
One realizes that a single integral yields the integrated density for (almost)
all wave function $\psi_{\nu}\left(  \xi\right)  $. The state $\psi_{0}\left(
\pi\xi\right)  $ lies roughly in the range $\xi<2^{2},$ i.e. the integrated
density $q_{\nu}\left(  \xi\right)  =\int_{0}^{\xi}\left\vert \psi_{0}\left(
\pi\xi\right)  \right\vert ^{2}d\xi^{\prime}$ increases in this range to 90\%.
Therefore the states $\psi_{\nu}\left(  \xi\right)  $ and $\psi_{N-1-\nu
}\left(  \pi\xi\right)  $ essentially lie in the range $\xi<2^{\nu+2}$ (in
units of $\lambda_{F}/2$). For a total of $N=50$ Wilson states the maximum
range of the wave functions is roughly $2^{N/2-1}=2^{26}$.

We may define as a ruler a linear array $I\left(  s\right)  $ where $s$ is an
integer$,$ $\left(  -N/2\leq s<N\right)  $ as%
\[
I\left(  s\right)  =2\int_{0}^{2^{s}}\frac{\sin^{2}\left(  \pi u\right)
}{\left(  \pi u\right)  ^{2}}du
\]
Then the integrated density of the state $\psi_{\nu}$ in the range from $0$ to
$2^{l}$ is given by
\[
q_{\nu}^{0}\left(  2^{l}\right)  =\int_{0}^{2^{l}}\left\vert \psi_{\nu}\left(
\xi^{\prime}\right)  \right\vert ^{2}d\xi^{\prime}=2\int_{0}^{\frac{2^{l}%
}{2^{\nu+2}}}\frac{\sin^{2}\left(  \pi u\right)  }{\left(  \pi u\right)  ^{2}%
}du=I\left(  l-\nu-2\right)
\]
$I\left(  l-\nu-2\right)  $ gives the integrated density of the wave function
$\psi_{\nu}$ within the radius $2^{l}$.

\subsubsection{Interference terms in the density}

The Wilson states $\psi_{\nu}\left(  \xi\right)  $ or $c_{\nu}^{\dag}$
represent the free electron states in the impurity problem. With the impurity
we express the ground state in terms of new states $a_{i}^{\dag}=%
{\textstyle\sum_{\nu=0}^{N-1}}
\alpha_{i}^{\nu}c_{\nu}^{\dag}$. Their integrated density is given by%
\[
\overline{\rho}_{i}\left(  2^{l}\right)  =\int_{0}^{2^{l}}\left\vert
{\textstyle\sum_{\nu=0}^{N-1}}
\alpha_{i}^{\nu}\psi_{\nu}\left(  \xi\right)  \right\vert ^{2}d\xi
\]

The quadratic terms can be evaluated with the same ruler $I\left(  s\right)  $
as before. But this time one has in addition interference terms $\psi_{\nu
}\left(  \xi\right)  \psi_{\nu+\lambda}\left(  \xi\right)  $. These terms
depend on two parameters, $\nu$ and $\lambda$. So one needs for each $\lambda$
a different ruler. Furthermore the interference terms depend on the sub-bands
of $\psi_{\nu}\left(  \xi\right)  $ and $\psi_{\nu+\lambda}\left(  \xi\right)
$. If both states lie either in the negative energy sub-band ($\nu,\nu
+\lambda<N/2$) or in the positive sub-band ($\nu,\nu-\lambda\geq N/2$) then
one obtains one set of rulers $I_{0}\left(  \nu,\lambda\right)  $ and if they
lie in opposite sub-bands then one obtains another set of rulers $I_{1}\left(
\nu,\lambda\right)  $. As an example one obtains%
\[
I_{0}\left(  s,\lambda\right)  =2\sqrt{2^{\lambda}}\int_{0}^{2^{s}}\frac
{\sin\left(  \pi u\right)  \sin\left(  \pi\frac{u}{2^{\lambda}}\right)
}{\left(  \pi u\right)  ^{2}}\cos\left(  3\pi u\left(  1-\frac{1}{2^{\lambda}%
}\right)  \right)  du
\]
For $I_{1}\left(  \nu,\lambda\right)  $ one has to replace the minus sign in
the cosine function by a plus sign. Furthermore one has to treat the terms
where $\nu+\lambda=N/2-1$ separately because one state lies at the Fermi level
and has a different cell width.

\subsubsection{The net integrated density}

If we occupy all Wilson states below the Fermi level then we obtain $%
{\textstyle\prod\limits_{\nu=0}^{n-1}}
c_{\nu}^{\dag}\Phi_{0}$ with $n=N/2$ and $\Phi_{0}$ the vacuum state. This
state is not really the free electron ground state. To obtain the latter we
have also to occupy the states $\overline{\varphi_{\kappa}\left(  \xi\right)
}$ and $\overline{\overline{\varphi_{\kappa}\left(  \xi\right)  }}$. They
don't interact with the impurity but they are occupied. Therefore we define as
quasi-vacuum $\Phi_{0}^{\prime}$ the state in which all states $\overline
{\varphi_{\kappa}\left(  \xi\right)  }$ and $\overline{\overline
{\varphi_{\kappa}\left(  \xi\right)  }}$ with $\kappa<1$ are occupied. Then
the ground state is $\Psi_{0}=$ $%
{\textstyle\prod\limits_{\nu=0}^{n-1}}
c_{\nu}^{\dag}\Phi_{0}^{\prime}$. This state has a constant electron density
in real space.

In the presence of the impurity the new ground state $\Psi_{new}=$ $%
{\textstyle\prod\limits_{i=0}^{n-1}}
a_{i}^{\dag}\Phi_{0}^{\prime}$ must also contain this quasi vacuum, i.e., the
non-interacting states must be occupied up to the Fermi level. Since the inert
states are occupied in $\Psi_{0}$ and $\Psi_{new}$ they cancel out when one
calculates the change in the electron density. The net density of the new
state $\Psi_{new}$ is the difference between $\rho\left(  \Psi_{new}\right)  $
and $\rho\left(  \Psi_{0}\right)  $. Since the inert states cancel out one can
ignore their existence during this calculation.

\end{document}